\documentclass[twocolumn]{pasj00}

\begin{document}

\SetVolumeData{}{57}{2}
\SetRunningHead{M. Murashima et al.}{X-Ray Spectra of Ton S180}
\Received{2003/06/24}
\Accepted{2004/12/29}

\title{X-Ray Spectra of the Narrow-Line Seyfert 1 Galaxy \\
Ton S180 in Comparison with Galactic Black Holes}

\author{Mio \textsc{Murashima}%
}
\affil{Department of Physics, The University of Tokyo,
       7-3-1, Hongo, Bunkyo-ku, Tokyo 113-0033, Japan}
\email{mio@amalthea.phys.s.u-tokyo.ac.jp}

\author{Aya \textsc{Kubota}}
\affil{Institute of Physical and Chemical Research,
       2-1, Hirosawa, Wako, Saitama 351-0198, Japan}

\author{Kazuo {\sc Makishima}}
\affil{Department of Physics, The University of Tokyo,
       7-3-1, Hongo, Bunkyo-ku, Tokyo 113-0033, Japan}
\affil{Institute of Physical and Chemical Research,
       2-1, Hirosawa, Wako, Saitama 351-0198, Japan}

\author{Motohide {\sc Kokubun}}
\affil{Department of Physics, The University of Tokyo,
       7-3-1, Hongo, Bunkyo-ku, Tokyo 113-0033, Japan}

\author{Soojing {\sc Hong}}
\affil{Department of Physics, Saitama University,
       255 Shimo-Okubo, Sakura-ku, Saitama 338-8570, Japan}
\affil{Institute of Physical and Chemical Research,
       2-1, Hirosawa, Wako, Saitama 351-0198, Japan}
\and
\author{Hitoshi {\sc Negoro}}
\affil{Department of Physics, Nihon University,
1-8-14, Kanda-Surugadai, Chiyoda-ku, Tokyo 101-8308, Japan}


\maketitle

\begin{abstract}
		
An analysis was made of 0.3--15 keV X-ray spectra of a Narrow-Line
Seyfert 1 Galaxy, Ton S180, using archival data from ASCA, RXTE, and
XMM-Newton.
At energies above 2.5 keV, a power-law with a photon index of
$\sim 2.3$ successfully and consistently reproduced the spectra from
all of these observatories.
Assuming this power-law component to extend toward
lower energies, a soft excess, which is one of the most remarkable
features of Narrow-Line Seyfert 1 Galaxies, is explained by another
power-law multiplied by a thermal cutoff at $\sim 0.4$ keV.  
Some similarities have been observed between this object and Galactic black
hole binaries in very high state, 
the latter being realized under high accretion rates. 
Attempts have been made to interpret the soft excess in terms of
Comptonization of the disk photons by an electron cloud surrounding
the accretion disk, like BHBs in a very high state.

\end{abstract}

\section{Introduction}

Narrow-Line Seyfert 1 galaxies (hereafter NLS1s) are identified by
the unusual narrownesses of their broad components of the H $\beta$ line;
the full width at half maximum is less than $2000$ km s$^{-1}$ 
(Osterbrock, Pogge 1985; V\'{e}ron-Cetty et al. 2001).
A general consensus is that NLS1s have relatively smaller black hole
masses than other more typical Seyfert 1 galaxies
(e.g. Wang, Lu 2001; Hayashida 2000; Kawaguchi 2003), together with
high accretion rates, even near the Eddington limit (e.g. Pounds et
al. 1995).

In the X-ray band, NLS1s show two remarkable features.
One is the X-ray spectrum, characterized by a steep power-law with a
photon index of 2 to 2.5 and a prominent soft excess below 2 keV
(e.g. Leighly 1999b). The soft excess is generally modeled
empirically by a blackbody with a temperature of 0.1 -- 0.2 keV.
The other is a large-amplitude variability on various time scales, from
several thousand seconds to more than several years (e.g. Leighly 1999a).
Although many attempts have been made to characterize this variability by
utilizing a cross-correlation function or other methods, no clear
correlations have been observed so far between the soft and hard
spectral components.

Of these X-ray properties of NLS1s, a particularly puzzling issue is
the nature of the soft X-ray component. Although we are tempted to
interpret it as being optically-thick thermal emission from accretion
disks, the observed temperature (0.1 -- 0.2 keV) is significantly higher than
the prediction from the standard accretion disk model (Shakura,
Sunyaev 1973) for the suggested black-hole mass of $10^{6-7}\ M_{\odot}$.
Furthermore, the strong variability exhibited by the soft component is
obviously distinct from the relative calmness of the optically thick
disk emission components from Galactic black hole binaries (hereafter
BHBs).

With regard to the high accretion rates of NLS1s, 
which appear to be one of their common physical properties,
we have revealed that 
BHBs under high accretion rates exhibit two characteristic states
(Kubota et al. 2001; Kobayashi et al. 2003; Kubota, Makishima 2004).
In one state, which is called a very high state,
the accretion-disk emission exhibits a significant hard
tail extending into higher energies, presumably due to Comptonization
of the original disk photons,
while in the other state the accretion disk deviates from the standard
model and becomes what is called a slim disk (Abramowicz et al. 1988;
Mineshige et al. 2000; Watarai et al. 2001). 
When we fit the spectrum of a BHB in either of  these states with the
standard disk model, the disk temperature is significantly higher
than a prediction of the standard model.
These two non-standard states may also explain (Watarai et al. 2001;
Kubota et al. 2002) the X-ray spectral properties of ultra-luminous
X-ray sources (ULXs ; Makishima et al. 2000), which are thought
to involve intermediate-mass black holes. 
We may therefore attempt to identify the soft excess of
NLS1s either with strongly Comptonized disk emission, 
or with the slim disk emission (Mineshige et al. 2000; Haba 2004).

Vaughan et al. (2002), based on XMM-Newton observations of the
prototypical NLS1 Tonantzintla (Ton) S180, located at a redshift of $z
\sim 0.062$, suggested
that the soft component prefers power-law to blackbody modeling, and
hence can be interpreted as arising via inverse Compton scattering.
This suggests that NLS1s are generally in a state analogous to the
very high state of BHBs, and Ton S180 is one of the best
candidates to examine this possibility.
In order to confirm this analogy between NLS1s and BHBs, we here focus on
this particular NLS1.

There have been a number of X-ray observations of Ton S180.
A long observation of Ton S180 was performed with ASCA in
1999 during a simultaneous campaign with repeated pointings with the
Rossi X-ray Timing Explorer (RXTE).
To the ASCA dataset acquired on this occasion, 
Romano et al. (2001), Turner et al. (2001),
and  Edelson et al. (2001) applied a series of standard
analysis, while focusing on the  spectral energy distributions, or
variability. In addition, this source had previously been observed in
the X-ray band by ROSAT (Fink et al. 1997), BeppoSAX
(Comastri et al. 1998),  and once by ASCA (Turner et
al. 1998). An observation with high-energy resolution was also
performed with the Chandra LETG (Turner et al. 2001). The most
recent observations were performed with XMM-Newton in 2000
December as mentioned above (Vaughan et al. 2002) and in 2002 June.
All of these
observations found typical features of NLS1s: a steep power-law
($\Gamma \sim 2.4$) and a featureless soft excess below 2 keV in the
spectrum, as well as rapid variability.

In the present work, we utilized three archival datasets out of these
observations.
One was the ASCA data in 1999, which still provide one of the
highest-quality spectra of Ton S180 at energies of 3 -- 10 keV
because of the very long exposure and the low background. 
We then utilized the RXTE data, which were simultaneous with ASCA
exposure.
After quantifying the hard-band spectral property using these ASCA 
and RXTE data,
we proceeded to an analysis of the X-ray spectra obtained by the first
XMM-Newton observation, in order to extract the soft excess.

We employed
a Hubble constant of $H_0 = 70$ km s$^{-1}$ Mpc$^{-1}$ and
a deceleration parameter of ${q_{_0}} = 0.5$ in the present analysis.

\section{Observation}

\begin{figure}[h]
\begin{center}
\rotatebox{270}{
\includegraphics[width=5.7cm]{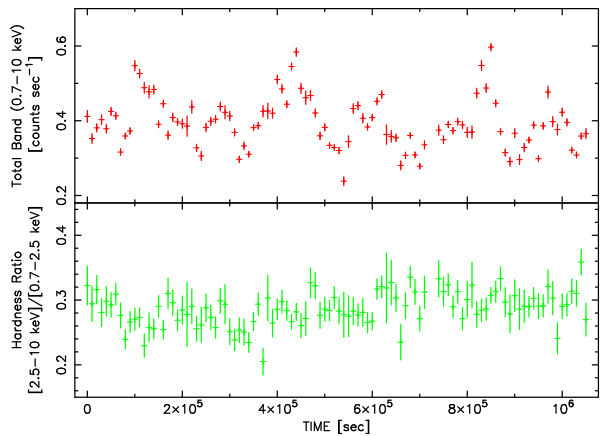}
}
\end{center}
\caption{(Top) Background-subtracted 0.7 -- 10 keV light curve of
Ton S180 observed with ASCA GIS 2$+$GIS 3, binned
into 10 ks. (Bottom) Time history of the 2.5 -- 10 keV to 0.7 -- 2.5 keV
hardness ratio.}
\label{lcurve}
\end{figure}

A long continuous observation of Ton S180 
was performed over 12 days with ASCA on 1999 December 3 -- 15, 
as a part of a multi-satellite, broad-band campaign (Turner et al. 2002).
After screening the data with the standard criteria, we obtained
a net exposure of 410 ks with the Gas Imaging Spectrometer (GIS :
Ohashi et al. 1996; Makishima et al. 1996)
and 356 ks with the Solid-State Imaging Spectrometer (SIS : Burke et
al. 1994).
Because of the long-term degradation of the SIS, 
we utilized mainly the GIS data in the present work,
and utilized the SIS data only to analyze spectra at energies above
2.5 keV.
In figure \ref{lcurve}, we show the 0.7 -- 10 keV light curve of Ton S180
obtained with GIS 2 and GIS 3,  
using a signal integration region of $3^{\prime}$ radius.
The typical count rate from the target is 0.4 c s$^{-1}$,
while that of background derived from off-source regions is 0.01 c
s$^{-1}$ when normalized to the same detector area;
we subtracted the background in advance.

The simultaneous RXTE observation was carried out from 1999 November
11 to December 16. 
During this period, RXTE regularly observed the target  once per
orbit, for 1 -- 2 ks each. 
We utilized data obtained with the Proportional Counter Array
(PCA) from December 3 to 15, covering the same
period as the ASCA observation. These datasets were 
acquired during 157 pointings, achieving a total exposure of 262 ks. 
The PCA covers an energy band of 2 to 60 keV (Jahoda et al. 1996) with
five proportional counter units (PCUs). 
Specifically, we utilized the standard-2 format data from PCU 0 and
PCU 2, because only these two out of the five PCUs were
active during these observations. 
Since we are interested in the average spectrum, we accumulated all of
these pointing data from the top layer of the two PCUs into a single
3 -- 15 keV spectrum, from which we subtracted the background spectrum
estimated using the latest mission-long faint model.
According to the data-analysis procedure recommended by the RXTE Guest
Observer Facility, the standard screening criteria were applied, to
achieve a net exposure of 48 ks.

The first observation of Ton S180 with XMM-Newton was performed in
December 2000. The obtained EPIC MOS 2 and PN data have a net
exposure of 29 ks and 21 ks, respectively.
The second observation was performed in June 2002, and 
the EPIC MOS2 and PN data were acquired for net exposures of 18 ks and
13 ks, respectively.  
All instruments were operated in a small-window mode with a medium
filter for both observations.
In order to avoid calibration uncertainties 
and degradation, which are especially prominent in a lower energy band,
such as a gain
problem of the PN small window mode around O-edge,
we utilized the data of only the first observation.
We extracted the MOS 2 and PN spectra of Ton S180 using
a signal integration region of $50^{\prime \prime}$ radius, 
and subtracted the background spectra obtained in an off-source region.

\begin{table*}[t]
\caption{Results of a single power-law fit to the spectra of Ton S180 at
energies above 2.5 keV.}
\label{hardtail}
\begin{center}
\begin{tabular}{lccc} \hline \hline
Instrument & Energy range [keV] \footnotemark[$*$]&
$\Gamma _{\rm hard}$ & $\chi ^2 / {\rm d.o.f.}$ \\ \hline
GIS 2+3	   & 2.5 -- 9  & $2.34 \pm 0.04$       & 165/225 \\
SIS 0	   & 2.5 -- 9  &$2.39 \pm 0.06$       & 132/141 \\
SIS 1	   & 2.5 -- 9  & $2.42 \pm 0.06$       & 127/119 \\
GIS 23 + SIS 0 + SIS 1 \footnotemark[$\dagger$]  & $2.5-9$  &
 $2.38 \pm 0.03$ & 427/486 \\
PCA	   & 3 -- 15  & $2.34 \pm 0.1$       & 36/24 \\
GIS 23 + SIS 0 + SIS 1 + PCA \footnotemark[$\dagger$]  & $2.5-15$  &
 $2.37 \pm 0.03$ & 453/507 \\
EPIC MOS 2  & 2.5 -- 10  & $2.25 \pm 0.10$       & 90/106 \\ 
EPIC PN	   & 2.5 -- 10  & $2.29 \pm 0.06$       & 218/224 \\ \hline
\multicolumn{4}{@{}l@{}}{\hbox to 0pt{\parbox{170mm}{\footnotesize
\vspace{0.3cm}
\par\noindent 
\footnotemark[$*$] Excluding 6 -- 7 keV.
\par\noindent 
\footnotemark[$\dagger$] A Combined fit to the spectra.
}\hss}}
\end{tabular}
\end{center}
\end{table*}

\section{Data Analysis and Results}

\subsection{Variability}

As can be seen in the GIS light curve (top  panel of figure
\ref{lcurve}), the 0.7 -- 10 keV signal intensity varied by a factor of
$\sim 3$ during the ASCA observation. The variation has a preferred
time scale of $\sim 1.5 \times 10^5$ s, although a detailed timing
analysis is beyond the scope of the present paper.
It is possibly the same phenomenon as the 2.08-day
($1.8 \times 10^5$ s) QPO that Halpern et al. (2003) found
in a long (33 days) EUVE observation, which partially overlaps
with the present ASCA exposure.

In the bottom panel of figure \ref{lcurve}, we also show the variation of
the hardness ratio, which is defined as the ratio of counts in 2.5 -- 10
keV to those in 0.7 -- 2.5 keV. The hardness ratio is not completely
constant during the observation,
implying some spectral variation.
However, the associated spectral change is rather small, as briefly
described in subsection 3.3.
We therefore analyzed mainly the time-averaged spectra.

\begin{table*}[t]
\caption{Results of conventional two-component model fits to the
$0.8-10$ keV ASCA GIS 2$+$GIS 3 spectrum of Ton S180. \footnotemark[$*$]
}
\label{asca-conventional}
\begin{center}
\begin{tabular}{lcccc} \hline \hline
Model & $kT$ [keV] & $\Gamma _{\rm soft}$  & $\Gamma _{\rm hard}$ &
$\chi ^2 / {\rm d.o.f.}$ \\ \hline
Blackbody + PL & $0.13 ^{+0.08}_{-0.07}$ & $-$ & $2.36 \pm 0.02$ &
330/393 \\
MCD + PL       & $0.16 \pm 0.01 $        & $-$ & $2.36 \pm 0.02$ &
328/393 \\
2PL            & $-$ & $5.22 ^{+0.40}_{-0.46}$ & $2.28 ^{+0.04}_{-0.05}$ & 
320/393 \\ \hline
\multicolumn{5}{@{}l@{}}{\hbox to 0pt{\parbox{170mm}{\footnotesize
\vspace{0.3cm}
\par\noindent 
\footnotemark[$*$] The absorption is fixed at the Galactic
line-of-sight value of $1.6 \times 10^{20}\ {\rm cm}^{-2}$.
}\hss}}
\end{tabular}
\end{center}
\end{table*}

\subsection{Time-Averaged Spectra}

\subsubsection{ASCA and RXTE spectra}

In order to extract the soft excess spectrum, we first need to
determine the spectral hard component as accurately as possible. 
We utilized the data obtained with GIS 2 and GIS 3 on-board
ASCA, because of its long exposure and high sensitivity
at higher energies. 
We fitted the time-averaged and background-subtracted GIS 2 $+$ GIS 3
spectrum with a single power-law model at energies above 2.5 keV,
but excluding 6 -- 7 keV where Fe-K lines are suggested.
As summarized in table \ref{hardtail}, we obtained an acceptable fit
with a photon index of the hard component as $\Gamma _{\rm hard} = 2.34
\pm 0.04$.
We also analyzed the SIS data from the same ASCA observation.
By fitting the background-subtracted SIS 0 and SIS 1 spectra separately
with a single power-law model
in the energy range above 2.5 keV, and further simultaneously fitting
the GIS and SIS spectra, we obtained the results summarized in Table
\ref{hardtail}. Thus, the $2.5-9$ keV ASCA spectra were successfully
and consistently fitted by a steep single power-law with $\Gamma_{\rm
  hard} \sim 2.4$.

In order to determine the hard component at higher energies,
we utilized the spectrum obtained with the PCA on-board RXTE.
A single power-law fit to the 3 -- 15 keV PCA spectrum was
successful, yielding a photon index of $\Gamma _{\rm hard} = 2.34 \pm
0.10 $, which is consistent with the results of ASCA.
Furthermore, by allowing relative normalizations to take free values,
we achieved a successful ($\chi ^2 / \nu \sim 453/507$) simultaneous
power-law fit to the spectra of ASCA and RXTE.
The obtained photon index, $\Gamma _{\rm hard} = 2.37 \pm 0.03$,
agrees with those from the individual instruments. 
We include these results in table \ref{hardtail}.

In figure \ref{hardtail}, the GIS spectrum shows an excess below 2 keV,
where the power-law model determined in $>$ 2.5 keV is extrapolated
toward lower energies employing the Galactic absorption column density
of $1.6 \times 10^{20} \ {\rm cm} ^{-2}$. 
We thus reconfirmed the soft excess that has been repeatedly observed
from this source.
Assuming again Galactic line-of sight absorption,
this soft excess is reproduced equally well by either a blackbody with
a temperature of 0.13 keV, or a multi color disk model (MCD :
\cite{mitsuda}) with a temperature of 0.16 keV, or a very steep
power-law with a photon index of 5.22. These results are summarized in
table \ref{asca-conventional}.

The hard power-law has a $2.5-10$ keV luminosity of $\sim 5 \times
10^{43}\ {\rm erg}\ {\rm s}^{-1}$,
while the absorption-corrected soft component has a $0.8-2.5$ keV
luminosity of $\sim 2 \times 10^{43}\ {\rm erg}\ {\rm s}^{-1}$ if
represented, e.g., by the MCD model.

\begin{figure}[t]
\begin{center}
\rotatebox{270}{
\includegraphics[width=5.7cm]{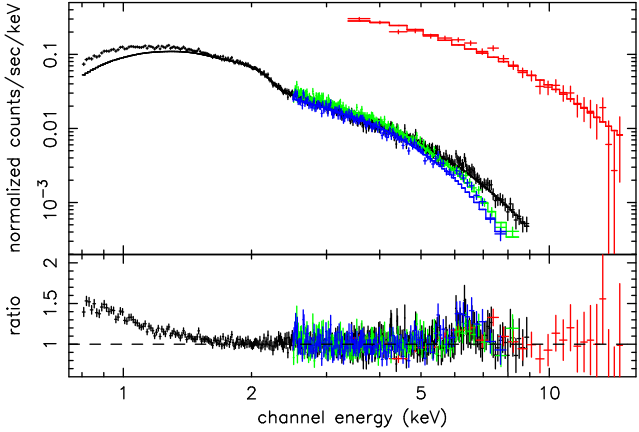}}
\end{center}
\caption{ASCA GIS (black), SIS 0 (green), SIS 1 (blue) and RXTE
PCA (red) spectra of Ton S180 fitted with a single power-law at
energies above 2.5 keV.
The top panel shows the data and the model with crosses
and solid lines, respectively. 
The bottom panel shows the ratio of the data to the model.
The fit is simultaneous among the four spectra.}
\label{asca-xte}
\end{figure}

\begin{figure}[t]
\begin{center}
\rotatebox{270}{
\includegraphics[width=5.7cm]{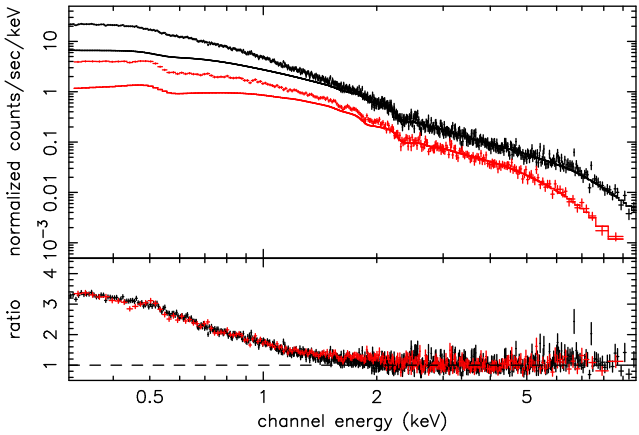}}
\end{center}
\caption{XMM EPIC PN (black) and MOS 2 (red) spectra of Ton S180
fitted with a power-law above 2.5 keV. 
The ratio of the data to the model is shown in the bottom panel.
The best-fit model is extrapolated toward lower energies to highlight
the soft excess.}
\label{xmm-hardtail}
\end{figure}

\begin{figure}[t]
\begin{center}
\rotatebox{270}{
\includegraphics[width=5.7cm]{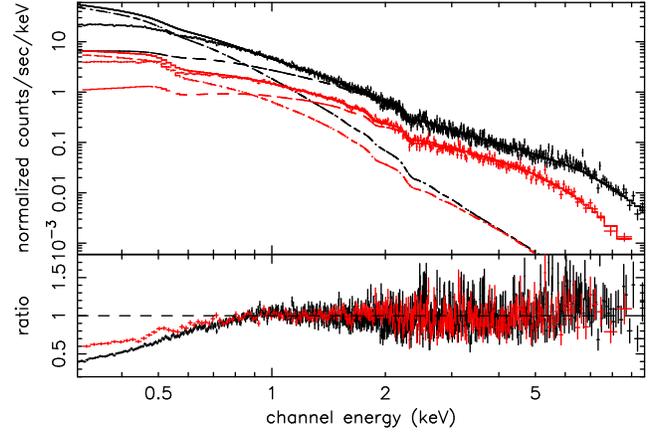}}
\end{center}
\caption{Same as figure \ref{xmm-hardtail}, but the fit was
performed with two power-law components over the $0.8-10$ keV range.
The best-fit model was extrapolated to energies below 0.8 keV.}
\label{xmm-2pl}
\end{figure}

\subsubsection{XMM-Newton spectra}

In order to better quantify the soft excess, we utilized the
XMM-Newton data, which extend down to an energy of 0.3 keV.
We first limited the energy range of analysis to 0.8 -- 10 keV
in order to examine the consistency with the ASCA results.
We determined the hard component in the same way as the ASCA analysis,
fitting the spectra of the EPIC MOS 2 and PN separately at energies above
2.5 keV with a single power-law. The obtained photon indices are
again summarized in table \ref{hardtail}.
Thus, the spectrum of Ton S180 at energies above 2.5 keV (up to 10 or
15 keV) is consistently described with a single power-law of $\Gamma
_{\rm hard} =$ 2.3 -- 2.4, by five instruments: the GIS and SIS
on-board ASCA, the RXTE PCA, and the two types of EPIC cameras
on-board XMM-Newton.

We fixed the obtained hard-component model, and added a second
steeper power-law to represent the soft excess.
This two-power-law model gave acceptable fits ($\chi ^2 / \nu \sim
864/808$) to the MOS 2 and PN spectra in the limited 0.8 -- 10 keV
range, with the photon indices for the second power-law of 4.3 and
4.8, respectively.
However, as shown in figure \ref{xmm-2pl}, the data fall significantly
below the model when we extrapolate the two-power-law model to
lower energies down to 0.3 keV. 
The fit remained unacceptable even if we allow the absorption to vary freely.

\begin{table*}[t]
\caption{Fits to the $0.3-10$ keV XMM-Newton spectra by a hard
power-law plus a power-law with an exponential cutoff. \footnotemark[$*$]}
\label{xmm-cutoff}
\begin{center}
\begin{tabular}{lcccc} \hline \hline
Instrument & $\Gamma _{\rm soft}$    & $E _{\rm cutoff}$  [keV] & 
$\Gamma _{\rm hard}$    & $\chi ^2 / {\rm d.o.f.}$ \\ \hline
MOS 2        & $2.44 ^{+0.41}_{-0.11}$ & $0.53 ^{+0.49}_{-0.20}$ &
$2.33 ^{+0.06}_{-0.18}$ & 274/260	\\ 
PN          & $2.10 ^{+0.03}_{-0.04}$ & $0.43 \pm 0.06$       &
$2.24 \pm 0.03$         & 686/675	\\
MOS 2 + PN \footnotemark[$\dagger$]
   & $2.09 \pm 0.03$         & $0.42 ^{+0.07}_{-0.05}$   &
$2.29 \pm 0.03$         & 974/934	\\ \hline 
\multicolumn{5}{@{}l@{}}{\hbox to 0pt{\parbox{170mm}{\footnotesize
\vspace{0.3cm}
\par\noindent 
\footnotemark[$*$] The absorption is fixed at the Galactic
line-of-sight value of $1.6 \times 10^{20}\ {\rm cm}^{-2}$.
\par\noindent 
\footnotemark[$\dagger$] A Combined fit to the spectra.
}\hss}}
\end{tabular}
\end{center}
\end{table*}

\begin{figure}[t]
\begin{center}
\rotatebox{270}{
\includegraphics[width=5.7cm]{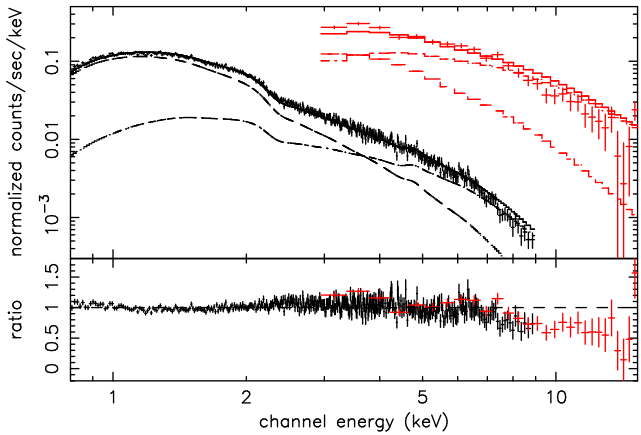}}
\end{center}
\caption{ASCA GIS (black) and RXTE PCA (red) spectra fitted
simultaneously by the two power-law model with indices fixed to  3.1
and 1.3, as determined by the XMM-Newton spectra.}
\label{asca-2pl-flat}
\end{figure}

\begin{figure}[t]
\begin{center}
\rotatebox{270}{
\includegraphics[width=5.7cm]{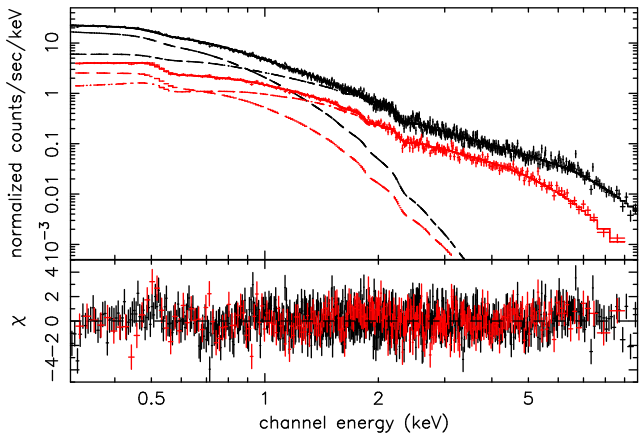}}
\end{center}
\caption{Same as figure \ref{xmm-hardtail}, but the fit was
performed over 0.3 -- 10 keV with the combination of a power-law and
another power-law with a thermal cutoff.
We show the reduced chi-squared results in the bottom panel to
demonstrate the acceptability of the fit.}
\label{xmm-cutoffpl}
\end{figure}

\begin{figure}[t]
\begin{center}
\rotatebox{90}{
\includegraphics[width=6cm]{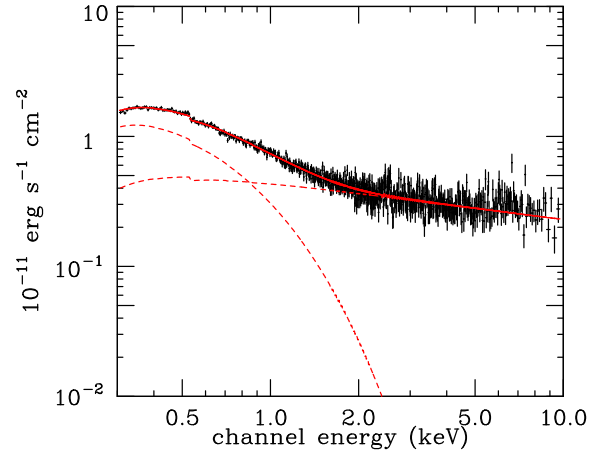}}
\end{center}
\caption{Unfolded PN spectrum (black) and model (red) 
in the form of $E \cdot F(E)$, where $F(E)$ is the energy flux density.
The data and the model are the same as those of figure \ref{xmm-cutoffpl}.}
\label{xmm-cutoffpl-flux}
\end{figure}

We then allowed the two power-law components to vary freely, to find
that the EPIC spectra could be explained reasonably well (with
$\chi ^2 / \nu = 1250/934$ ) by the sum of two power-laws, with photon
indices of 3.1 and 1.3, both absorbed by the Galactic column.
This is consistent with Vaughan et al. (2002), who reported that the
overall ($0.3-10$ keV) XMM-Newton spectra are reproduced by a very
flat hard tail with a photon index of 1.5, together with a steep
($\Gamma = 3.1$) power-law, which accounts for the softer part of the
EPIC spectra. However, this result is inconsistent with that of ASCA
and RXTE derived in sub subsection 3.2.1.
Actually, as shown in figure \ref{asca-2pl-flat}, the two-power-law
model with the indices fixed at 3.1 and 1.3 (but with their
normalizations both left free) gives an inadequate fit to the ASCA GIS
data, which have higher statistics than the XMM-Newton data at higher
energies. As we extrapolate the same model to a higher band up to 15
keV to include the RXTE spectrum, the model further deviates from the
observation  (with $\chi ^2 / \nu = 641/422$ ). 
Thus, the two-power-law model determined with the XMM-Newton spectra
fails to explain those of ASCA and RXTE toward higher energies.

\subsubsection{Unified spectral modeling}

The ASCA and RXTE spectra covering 0.8 -- 15 keV were simultaneously
reproduced by a two-power-law model with photon indices of 5.22 and
2.28. The 0.3 -- 10 keV XMM-Newton spectra were also explained by
another two-power-law model with smaller photon indices of 3.1 and
1.3. These two models are hardly distinguished in the energy band of
0.8 -- 9 keV.
Therefore, the spectra of ASCA and XMM-Newton are basically similar in
energies where their sensitivities significantly overlap, in spite of
the difference in the epoch of observation.
Nevertheless,
the model of ASCA and RXTE cannot reproduce the XMM-Newton spectra
at lower energies below 0.8 keV, and that of XMM-Newton cannot explain
the ASCA and RXTE spectra at higher energies above $\sim 8$ keV. 
In short, both two-power-law models cannot explain the overall X-ray
spectrum of Ton S180.
We should thus adopt the hard component determined by the ASCA and
RXTE data, and utilize the XMM-Newton data to better constrain the
soft excess.
A model that can explain a whole spectrum of this NLS1 will
therefore consist of a power-law with an index of $\sim 2.3$,
and an appropriate component for the soft excess.
As suggested by figure 3, the latter must have a mildly convex shape,
with a steep slope from $\sim 2$ keV down to $\sim 0.5$ keV, but a
gradual flattening to lower energies.

We adopted a power-law multiplied by a thermal (exponential) cutoff
factor to reproduce the curvature of the soft excess.
All of the parameters of this component and of the hard power-law were
left free to vary, except for the absorption, which was again fixed to the
Galactic value. As shown in figure \ref{xmm-cutoffpl} and table
\ref{xmm-cutoff}, this model gave acceptable fits to the MOS 2 and
PN spectra, either separately or
jointly, over the entire $0.3-10$ keV energy range.
The obtained joint fit ($\chi ^2 / \nu = 974/934$) is actually
considerably better than the two-power-law fit ($\chi ^2 / \nu =
1250/934$) that we obtained in sub subsection 3.2.2.  
The slope of the hard component has remained close to what we 
determined in sub subsection 3.2.1 and 3.2.2, while ignoring the data
below 2.5 keV.
The index of the soft excess is similar to that of the hard component,
and the cutoff energy is $\sim 0.4$ keV, in agreement with the visual
impression from the spectral ratio in figure 3.

Thus, we have successfully reproduce the overall XMM-Newton spectra of
Ton S180 by combining the $\Gamma \sim 2.3$ power-law and another
power-law with an exponential cutoff, representing the hard component
and the soft excess, respectively.
In figure \ref{xmm-cutoffpl-flux}, we show the unfolded PN spectrum 
and the best-fit model in the form of $E \cdot F(E)$,
where $F(E)$ is the energy flux density. 
This model implies a hard-component luminosity of 
$\sim 3 \times 10^{43}\ {\rm erg}\ {\rm s}^{-1}$ and a
soft-excess luminosity of $\sim 2 \times 10^{43}\ {\rm erg}\ {\rm s}^{-1}$,
in the energy ranges of $2.5-10$ keV and $0.8-2.5$ keV, respectively.

\subsection{Spectral Variability}

In order to briefly investigate the spectral variability,
we divided the 12 days of long ASCA data into 12 subsets of $\sim35$ ks
each, and fitted with the individual spectra using the MCD and
power-law model, which is one of the conventional models.
The obtained disk temperatures are almost constant, 0.16 keV, within
an error of 0.06 keV. 
The indices slightly change from $\sim 2.3$ to $\sim 2.5$ with
relatively large errors of $\sim 0.1$.  
The luminosity of the power-law varies together with the total
luminosity by a factor of 1.4,
and its ratio to the MCD luminosity changes between $\sim 4$
and $\sim 7$. Due to large errors compared to the range of parameter
changes, we found no significant correlations among the model
parameters, such as a positive correlation between the disk
luminosity and temperature that is reported in the case of BHB
(Tanaka, Lewin 1995; Tanaka 1997; and references therein).

\section{Discussion}

The standard accretion model predicts a disk temperature of $\sim 36$ eV 
for a black hole with a mass of $\sim 1.2
\times 10^7 M_{\odot}$ (Wang, Lu 2001), assuming a face-on geometry, 
a color hardening factor of $\kappa = 1.7$  (a ratio of a color
temperature to an effective one; Shimura, Takahara 1995), and a luminosity
close to the Eddington limit;
the latter two assumptions yield the maximum disk temperature.
This theoretical model also predicts a disk inner radius of 
$\sim 3 \times 10^7$ km.
With the ASCA spectra,
we obtained the temperatures of the soft
excess as 0.13 keV and 0.16 keV by fitting with the blackbody and the
MCD, respectively, as listed in table \ref{asca-conventional}.
We also obtained a disk inner radius of $\sim 1 \times 10^6$ km
using the disk temperature and luminosity.
As Vaughan et al. (2002)  argued based on the XMM-Newton spectra, 
these conventional fits give a four-times higher temperature
and a much smaller inner disk radius than is expected.
Consequently, the soft excess cannot be interpreted as being optically
thick thermal emission directly from a standard accretion disk.

Although an XMM-Newton observation was performed one year after
that with the other two satellites, the source was found in very
similar states on these two occasions; in fact, the luminosity
agrees within 30\%, and the spectral shape is very similar.
Our analysis using the five instruments on-board the three satellites
have consistently shown that the 2.5 -- 15 keV spectra of Ton S180 are
reproduced by a single power-law flux with $\Gamma _{\rm hard} \sim 2.3$.
Assuming that this single power-law flux continues toward energies below
2.5 keV, we have found that the soft excess has a mildly convex shape.
The slope of the soft excess toward lower energies is estimated to be
$\sim 2.1$, and its curvature toward higher energies is reproduced
by an exponential factor of $\sim 0.4$ keV.

As described in section 1, a possible interpretation of the soft excess
is the slim-disk scenario. In fact, this interpretation has been
proposed both observationally (Haba 2004) and theoretically (Mineshige
et al. 2000; Kawaguchi 2003; R{\' o}{\. z}a{\' n}ska et al. 2004).
In the present work, we employed the very high-state scenario
as an alternative working hypothesis to explain NLS1s.
As a result, we found several interesting similarities
between Ton S180 and BHBs in this particular state.
Firstly, the power-law shaped hard component
with a relatively steep photon index of $\sim 2.3$,
which dominates the 0.3 -- 15 keV spectrum of Ton S180,
is very reminiscent of the characteristic spectral slope
observed from BHBs in the very high state
over a few to a few tens of keV range (Kubota, Makishima 2004).
Secondly, the rapid variability characterizing NLS1s
is also commonly observed from BHBs in the very high state
(Miyamoto et al. 1993; McClintock, Remillard 2003),
while not being confirmed so far from BHBs in the slim-disk state.
Finally, the intensity variation of a BHB in a very high state
often takes the form of quasi-periodic oscillations,
with a frequency of 1 -- 20 Hz (McClintock, Remillard 2003).
If we scale this frequency by 6 orders of magnitude
with the difference in the black-hole mass between BHBs and Ton S180,
we expect to observe a similar behavior from Ton S180
with a frequency of $10^{-5}$ Hz, or a period of $\sim 1 \times 10^5$ s.
The variation that we observe in figure 1
is indeed suggestive of such oscillations.

If our analogy between Ton S180 and very-high-state BHBs is correct,
and the emission from the latter is indeed dominated by Comptonized
disk photons, and the soft excess from Ton S180 is also inferred to be
arising via Comptonization, as already suggested by Vaughan et al. (2002).
The shape of the soft excess that we have determined,
a power-law multiplied by an exponential factor,
is in a qualitative agreement with a prediction by
the theory of unsaturated Comptonization (Sunyaev, Titarchuk 1980).
The seed photons may be supplied by an optically-thick accretion disk,
which may be visible at optical -- ultraviolet frequencies (Turner
et al. 2002).
The scattering electron cloud is suggested
to have a temperature of $\sim 0.4$ keV;
this is significantly higher than the disk temperature of 36 eV,
which is predicted by the standard disk scenario,
and suggested by the optical -- ultraviolet data (Turner et al. 2002).
However, this electron temperature is much lower than is usually
invoked for BHBs, and hence requires rather large optical depths
to produce sufficient Comptonization effects.
Magdziarz et al. (1998) fitted the soft excess in NGC 5548,
which is a typical Seyfert 1 galaxy, by Comptonization due to such 
a cool ($kT \sim 0.3$ keV) and optically thick ($\tau \sim 30$) plasma.
The reality and physical details of electron scattering
with such a low electron temperature and large optical depths
are a subject of future study.

In summary, we have accurately quantified the $0.3-15$ keV
spectrum of Ton S180 using the three satellites,
and pointed out empirical similarities
between this NLS1 and BHBs in very high states.
This analogy suggests
that the soft excess of Ton S180
arises via Comptonization of the disk photons,
by an electron cloud surrounding the accretion disk.


\end{document}